\documentclass[twocolumn,showpacs,prl,preprintnumbers,aps]{revtex4}
\usepackage{dcolumn}% Align table columns on decimal point
\usepackage{epsf}

\long\def \blockcomment #1\endcomment{}

\def\cp{{\cal P}}

\def\svev#1{\left\langle #1\right\rangle}       % variable < >

\def\Im{{\rm Im\,}}
\def\tr{{\rm tr}\,}
\def\lsupp{l_{s}}
\def\ltail{l_{l}}
\def\mres{m_{\text{res}}}
\def\bj{\bar{\psi}}
\def\tG{\tilde{\Gamma}}
\def\tH{\tilde{H}}
\def\tl{\tilde{l}}
\def\tmu{\tilde{\mu}}

\def\PRD#1{Phys.\ Rev.\ D {\bf #1}}
\def\PRP#1{Phys.\ Rep.\ {\bf #1}}
\def\APH#1{Ann.\ Phys.\ (NY) {\bf #1}}

\def\NPB#1{Nucl.\ Phys.\ {\bf B#1}}

\begin{document}

\title{Mobility edge in lattice QCD}

\author{Maarten Golterman}%
%\email{maarten@stars.sfsu.edu}
 \affiliation{Department of Physics and Astronomy,
San Francisco State University, San Francisco, CA 94132, USA}

\author{Yigal Shamir}
%\email{shamir@post.tau.ac.il}
\author{Benjamin Svetitsky}
%\email{bqs@julian.tau.ac.il}
 \affiliation{School of Physics and Astronomy, Raymond and Beverly
Sackler Faculty of Exact Sciences, Tel~Aviv University, 69978
Tel~Aviv, Israel}

\begin{abstract}
We determine the location $\lambda_c$ of the mobility edge in the spectrum
of the hermitian Wilson operator on quenched ensembles.
We confirm a theoretical picture of localization proposed
for the Aoki phase diagram.
When $\lambda_c>0$ we also determine some
key properties of the localized eigenmodes
with eigenvalues $|\lambda|<\lambda_c$.
Our results lead to simple tests for the validity of simulations with
overlap and domain-wall fermions.
\end{abstract}


\maketitle

Localization of electronic wave functions is a familiar phenomenon
in disordered systems \cite{DJT}. Recently we conjectured
\cite{lcl} that a similar phenomenon takes place in lattice QCD
with Wilson fermions: In an ensemble of gauge configurations,
the low-lying eigenmodes of the hermitian Wilson operator can be
localized, up to some mobility edge
above which they become extended.  This observation has two
important applications.  First, it helps resolve a paradox
in the quenched theory with negative bare mass $m_0$,
where simulations with two valence quarks
have discovered a condensate that breaks the
isospin symmetry in regions without Goldstone bosons \cite{scri}.
Second, there are important implications for large-scale simulations of
QCD with domain-wall \cite{dwf} and overlap \cite{overlap} fermions.
Both of these formulations
are based on Wilson fermions with negative $m_0$. Quenched
as well as unquenched calculations with these fermions
will thus be sensitive to the spectrum of the Wilson operator.
It turns out that an understanding of the localization properties
is important for ensuring chirality and locality.

In two-flavor QCD with Wilson fermions,
part of  the ``supercritical'' region ($-8<m_0<0$)
is the so-called Aoki phase \cite{aoki}, where a pion condensate
$\langle\pi_3\rangle$ breaks both parity and isospin symmetry.
Inside the Aoki phase one pion is massive, whereas the other two
pions are Goldstone bosons. Outside the Aoki phase (e.g., for weak
coupling, away from the critical values $m_0=0,$ $-2,$ $\ldots$) all
pions are massive.

It is this supercritical massive phase that presents the conundrum
in the quenched theory. The word ``quenched'' here
can refer to any ensemble of gauge configurations generated
without the Wilson-fermion determinant, whether
a pure gauge ensemble or an ensemble with dynamical fermions of some other
type, such as domain-wall fermions
\footnote{
  For a chiral lagrangian study of the pure gauge ensemble
  with Wilson valence quarks, see
  M.~Golterman, S.~Sharpe and R.~Singleton, contribution to Lattice 2004.
}.
Both the massless and the massive supercritical phases
support a non-zero density of near-zero eigenmodes, and
hence a condensate through the Banks--Casher relation
$\svev{\pi_3}=2\pi\rho(0)$ \cite{BC}, where $\rho(\lambda)$ is the spectral density of
the hermitian Wilson operator $H_W$.
Why are there no Goldstone bosons?
In Ref.~\onlinecite{lcl} we proposed a detailed
physical picture
which resolves this puzzle. A key observation (first made
in Ref.~\onlinecite{ms}) is that, in a quenched system, localization provides
an alternative to the Goldstone theorem. In this Letter we present
numerical evidence supporting and illustrating this picture.
The implications for domain-wall and overlap fermions are
thus made more concrete.

In order to develop this physical picture, we probe our quenched
ensemble with the two-flavor fermion action
\begin{eqnarray}
  S_F &=& \bj (D - (\lambda +i\tau_3 m_1)\gamma_5) \psi
\nonumber\\
  &=&  \bj (H_W - (\lambda +i\tau_3 m_1)) \psi',
\label{S}
\end{eqnarray}
where $\psi'=\gamma_5\psi$, and
$\tau_k$ are Pauli matrices acting in flavor space.
$H_W=D\gamma_5$ is hermitian.  Explicitly,
\begin{equation}
   D =  \left(\begin{array}{cc}
      W + m_0    & -C     \\
      C^\dagger  & W + m_0
       \end{array}\right),
\label{DW}
\end{equation}
where
$C_{xy} = {1\over 2} \sum_{\mu=1}^{4} \left(\delta_{x+\hat\mu,y} U_{x\mu}
    - \delta_{x-\hat\mu,y} U^\dagger_{y\mu} \right) \sigma_\mu$
and
$W_{xy} = 4\delta_{xy} -{1\over 2} \sum_{\mu=1}^{4} \left(\delta_{x+\hat\mu,y} U_{x\mu}
               + \delta_{x-\hat\mu,y} U^\dagger_{y\mu} \right)$.
Each entry is a $2\times 2$ matrix, with $\sigma_\mu=(\vec\sigma,i)$, and
$\sigma_k$ three Pauli spin matrices. The link variables $U_{x\mu} \in SU(3)$
constitute the random field in which the
fermions move. The parameter $\lambda$ will allow us to study the
spectral density $\rho(\lambda)$ of $H_W$ via a condensate.
$m_1$ is  a ``twisted mass'' which breaks isospin \cite{aoki,tm},
and acts as an external magnetic field for the condensate of interest.
Neither $\lambda$ nor $m_1$ appears in the Boltzmann weight of the ensemble.

For any $\lambda$ and $m_1$ one derives the Ward identity
\begin{equation}
  \sum_\mu \partial^*_\mu \svev{J^+_\mu(x)\,\pi^-(y)}
  + 2 m_1 \svev{\pi^+(x)\,\pi^-(y)}
    \! = \! \delta_{xy} \svev{\pi_3(y)}.
\label{WI}
\end{equation}
Here $\partial^*_\mu f(x) = f(x) - f(x-\hat\mu)$ and
$J^+_\mu(x)$ is the flavor-changing vector current
\footnote{$J^+_\mu(x)$ is conserved for $m_1=0$ in the unquenched
  theory; the quenched theory is ill defined for $m_1=0$ (see
  Ref.~\onlinecite{lcl}).
},
and $\pi^\pm=i\bj\gamma_5 \tau_\pm\psi$ and $\pi_3=i\bj\gamma_5 \tau_3\psi$,
with $\tau_\pm=(\tau_1\pm i\tau_2)/2$.
Introducing the Green function  $G = (H_W - \lambda - im_1)^{-1}$
one has
\begin{equation}
  \svev{\pi_3} = (2/V_4)\, \Im \tr \svev{G},
\label{pi3}
\end{equation}
where $V_4$ is the four-volume.
This implies a generalized Banks--Casher relation
\begin{equation}
  \lim_{m_1\to 0} \svev{\pi_3} = 2\pi\rho(\lambda) .
\label{rho}
\end{equation}
Thus the spectral density $\rho(\lambda)$ is an
order parameter for flavor symmetry breaking in the quenched theory
with fermion action~(\ref{S}).
The easiest way to calculate $\rho(\lambda)$ is in fact through
Eqs.~(\ref{pi3}) and~(\ref{rho}).

The two-point function $\Gamma(x,y;\lambda)=\svev{\pi^+(x)\,\pi^-(y)}$
represents correlations of the eigenmode densities of $H_W$. This
is readily seen from its spectral decomposition,
\begin{eqnarray}
  \Gamma(x,y;\lambda) &=&
  \left\langle
    \sum_{kn}
    \Psi_n^\dagger(x) \Psi_k(x) {1\over \lambda_k - \lambda + im_1}\right.
\nonumber\\
    &&\left.\times\Psi_k^\dagger(y) \Psi_n(y) {1\over \lambda_n - \lambda - im_1}
  \right\rangle,
\label{dblsum}
\end{eqnarray}
where $\Psi_n$ is the eigenmode of $H_W$ with eigenvalue
$\lambda_n$. We calculate it at zero three-momentum,
$\Gamma(t;\lambda) =(\pi V_3)^{-1}
\sum_{\vec{x}\vec{y}}\Gamma(0,\vec{x},t,\vec{y};\lambda)$,
where $V_3$ is the spatial volume. The mobility edge $\lambda_c$ is determined
as the value of $\lambda$ where these correlations become
long-ranged as $m_1\to 0$, that is, when the large-$t$
behavior of $\Gamma(t;\lambda)$ changes from exponential
($0\leq |\lambda|<\lambda_c$) to power law ($|\lambda|>\lambda_c$)
in this limit. Above $\lambda_c$ one has extended
modes, and the long-range density--density correlations play the
role of Goldstone bosons for flavor symmetry breaking.  Below
$\lambda_c$ there are no long-range correlations, and {\it no}
massless pole in $\svev{J^+_\mu(x)\,\pi^-(y)}$.
How, then, can the Ward identity~(\ref{WI}) be satisfied in the limit
$m_1\to 0$? The answer is that, when $\rho(\lambda)$ arises from
exponentially localized modes, the quenched two-point function
$\Gamma(x,y;\lambda)$ diverges as $1/m_1$
in this limit \cite{ms,lcl}. In fact
\footnote{This is rigorously true in finite volumes.  It is
  {\em not} true  in the $V_3\to\infty$ limit
  if $\rho(\lambda)$ contains contributions of extended modes.
},
\begin{eqnarray}
  \Gamma(x,y;\lambda)\!\!\! &=& \!\!\!{1\over m_1}
  \left\langle
  \sum_n |\Psi_n(x)|^2 |\Psi_n(y)|^2
  {m_1 \over (\lambda_n - \lambda)^2 + m_1^2}
  \right\rangle
\nonumber\\
  &&\quad + O(1).
\label{limm1}
\end{eqnarray}
As $m_1\to0$, the expectation value in Eq.~(\ref{limm1}) is
non-zero if and only if $\rho(\lambda)\ne 0$. It thus provides
a mechanism for saturating the Ward identity without Goldstone
bosons.

If the mobility edge is at $\lambda=0$, Goldstone
bosons dominate the correlation function; hence we may
take $\lambda_c=0$ to be the \textit{definition}
of the Aoki phase.

We have determined the value of $\lambda_c$ at several locations
in the $(\beta,m_0)$ plane. For each value of $\beta$ we generated an
ensemble of 120 quenched configurations using the standard
plaquette action. The four-volume was $16^4$, with periodic boundary
conditions for all fields. Our measurements were mostly
done at $m_0=-1.5$ [hopping parameter $\kappa=(8+2m_0)^{-1}=0.2$],
which is roughly the value used in domain-wall and overlap
simulations. We measured $\Gamma(t;\lambda)$ using random
sources on time slices 0 and $t$. We extracted a mass
$M=M(\lambda,m_1)$ from $\Gamma(t;\lambda)$ at $m_1$ values between 0.01
and~0.07, and extrapolated to $m_1= 0$ by fitting to
$M^2 = \mu^2(\lambda) + \alpha(\lambda)\, m_1$
\footnote{
  This fit usually works well.
  Above the mobility edge $M$ is the mass of a pseudo-Goldstone boson
  and should scale roughly as $M^2 \propto m_1$.
  Negative extrapolated values may be a signal of chiral logs and/or
  finite-volume effects.
}.
The results for $\beta=5.85$ are shown in the second column of
Table~\ref{table1}. One sees that $\mu^2$ starts falling rapidly
above $\lambda=0.1$.

%%%%%%%%%%%%
\begin{table}[t]
\caption{%\label{tab:table1}
Exponential falloff rate of the density--density correlator
$\Gamma(t;\lambda)$ at $\beta=5.85$, $m_0=-1.5$, for $m_1\to 0$. Errors are
statistical only. $\mu$ is the extrapolation of masses determined
at $m_1\neq0$, while $\tmu$ is extracted by extrapolation of
$m_1\Gamma(t;\lambda)$ to $m_1=0$.
\label{table1}}.
\begin{ruledtabular}
\begin{tabular}{ddc}
\lambda & \mu^2 & $\tmu^2$\\
\hline
0.0 &  2.42(6) & 1.9--2.3 \\
0.1 &  1.99(8) & 1.3--1.9 \\
0.2 &  1.18(5) & 0.9--1.1 \\
0.3 &  0.21(3) & 0.4--0.5 \\
0.4 & -0.04(2) & 0.12 \\
0.5 & -0.04(2) & 0.05 \\
0.6 & -0.05(2) & 0.01 \\
\end{tabular}
\end{ruledtabular}
\end{table}
%%%%%%%%%%%%%

By definition $\mu^2(\lambda)$ drops to zero at the mobility edge $\lambda_c$.
We determine $\lambda_c$ by linear
extrapolation from the last two points with positive $\mu^2$.
Our results are compiled in
Table~\ref{table2}. Consider first the $m_0=-1.5$ results. For
reference, we include the free-theory limit ($\beta=\infty$)
\cite{lcl}, where $\lambda_c$ coincides with the
gap of the free $H_W$. At $\beta=6.0$, $\lambda_c$ is still close to its
free-field value. The curve $\lambda_c(\beta)$ steepens before reaching
zero somewhere between $\beta=5.6$ and $\beta=5.5$, where we enter
the Aoki phase.

%%%%%%%%%%%%%
\begin{table}[t]
\caption{%\label{tab:table2}
Mobility edge $\lambda_c(\beta,m_0)$ and  (when $\lambda_c>0$) interpolated spectral density
$\rho(\lambda_c)$.
Where no error is shown, the
(statistical) error is less than one in the last digit.
\label{table2}}
\begin{ruledtabular}
\begin{tabular}{dddc}
\beta & m_0 &  \lambda_c & $\rho(\lambda_c)$ \\
\hline
\infty    &  -1.5 &  0.5      &       \\
6.0       &  -1.5 &  0.41     & 0.14  \\
5.85      &  -1.5 &  0.32     & 0.08  \\
5.7       &  -1.5 &  0.25     & 0.07  \\
5.6       &  -1.5 &  0.14(2)  & 0.05  \\
5.5       &  -1.5 &  0.0      & --    \\
5.4       &  -1.5 &  0.0      & --    \\
\hline
5.7       &  -2.0 &  0.21     & 0.14  \\
5.7       &  -2.4 &  0.03(2)  & 0.06  \\
\end{tabular}
\end{ruledtabular}
\end{table}
%%%%%%%%%%%%

Table~\ref{table2} also shows results at two other $m_0$ values for $\beta=5.7$.
The $m_0=-2.4$ result suggests that one is near the boundary of the
Aoki phase \footnote{
  This is one of the ``fingers'' of the Aoki phase
  in the quenched theory. The fingers may not exist in the
  theory with dynamical Wilson fermions. See
  E.~M.~Ilgenfritz {\it et al.},
%``A numerical reinvestigation of the Aoki phase with N(f) = 2 Wilson fermions
%at zero temperature,''
  Phys.\ Rev.\ D {\bf 69}, 074511 (2004);
%[arXiv:hep-lat/0309057].
%%CITATION = HEP-LAT 0309057;%%
  F.~Farchioni {\it et al.},
%``Twisted mass quarks and the phase structure of lattice QCD,''
  arXiv:hep-lat/0406039.
%%CITATION = HEP-LAT 0406039;%%
}.
We find only a small change between $m_0=-2.0$ and
$m_0=-1.5$, consistent with the finding of
Ref.~\onlinecite{scri} that the spectral properties of $H_W$ vary slowly
over this range. This is why we have explored mainly the
$\beta$-dependence at fixed $m_0=-1.5$.

%>>>>>>>>>>>>>>>>%
% FIGURE         %
%>>>>>>>>>>>>>>>>%
\begin{figure}[b]
%\vspace*{0.2cm}
\centerline{
\epsfxsize=6.0cm
\epsfbox{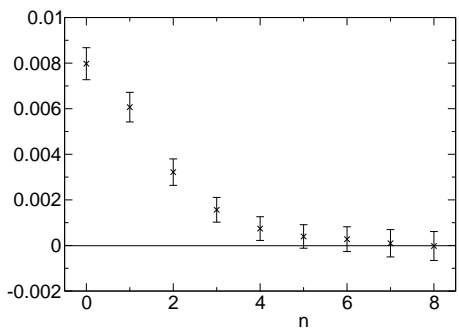}
}
%\vspace*{0.4cm}
%\begin{quotation}
\caption{Coefficient of the $1/m_1$-divergence in $\tG(\omega_n;\lambda)$
for $\beta=5.7$, $m_0=-1.5$, $\lambda=0.0$. \label{fig1}}
%\end{quotation}
%\label{fig}
%\vspace*{0.5cm}
\end{figure}
%<<<<<<<<<<<<<<<<%
% FIGend         %
%<<<<<<<<<<<<<<<<%

In order to account \cite{lcl} for the absence of a massless pole
in Eq.~(\ref{WI}), the $1/m_1$-divergence in
$\Gamma(x,y;\lambda)$ must persist for a range of momenta $p$,
and its coefficient should depend
smoothly on $p$. To confirm this, we calculated the Fourier
transform $\tG(\omega_n;\lambda) = \sum_t \cos(\omega_n t) \Gamma(t;\lambda)$,
where $\omega_n=2\pi n/16$, and extrapolated $m_1 \tG(\omega_n;\lambda)$
linearly to $m_1=0$
\footnote{
The factor
  $m_1/[(\lambda_n-\lambda)^2+m_1^2]$
  in Eq.~(\ref{limm1}) justifies the linear fit.
  The contribution of modes with
  $|\lambda_n-\lambda| \sim m_1$ to $m_1 \Gamma(t;\lambda)$
  is roughly constant, while that of ``bulk'' modes with
  $|\lambda_n-\lambda| \gg m_1$ vanishes linearly with $m_1$.
}.
Results are shown in
Fig.~\ref{fig1} for $\lambda=0.0$ at $(\beta,m_0)=(5.7,-1.5)$. The
$\omega$-dependence of the $1/m_1$-divergence is indeed smooth.
For comparison, we repeated the calculation for $\lambda=0.5$, which is
above $\lambda_c$. The extrapolation of $m_1\tG(0;\lambda)$ to $m_1=0$ is
straightforward, as it must be since this gives $\rho(\lambda)$ according
to Eqs.~(\ref{WI}) and~(\ref{rho}).
Doing the same with
\mbox{$\tG(\omega_n\neq0;\lambda)$}, however, leads to a
huge $\chi^2$.
This confirms the
qualitative difference between $|\lambda|<\lambda_c$ and $|\lambda|>\lambda_c$.

We present in Table~\ref{table3} some quantities that further illustrate
properties of the localized modes, for the same $(\beta,m_0)$ as in
Table~\ref{table1}.
Using them, we can address the question of whether below the mobility edge
$\rho(\lambda)$ arises from well-separated, exponentially
localized eigenmodes.

%%%%%%%%%%%%%
\begin{table}[t]
\caption{%\label{tab:table3}
Spectral properties for $\beta=5.85$, $m_0=-1.5$. The mobility edge
is at $\lambda_c=0.32$, marked by the horizontal line in the table.
\label{table3}}
\begin{ruledtabular}
\begin{tabular}{dlddd}
\lambda & $\rho(\lambda)$ & R & \lsupp & \ltail \\
\hline
0.0 &  0.0011(1)  &  17.   &  3.1(4)   &  0.64(1)  \\
0.1 &  0.0019(1)  &  15.   &  3.5(5)   &  0.71(1)  \\
0.2 &  0.0088(4)  &  10.   &  3.7(4)   &  0.92(2)  \\
0.3 &  0.056(1)   &  6.5   &  4.3(3)   &  2.2(2)   \\
\hline
0.4 &  0.168(7)   &  4.9   &   4.7(8)  &  -         \\
0.5 &  0.27(1)    &  4.4   &  11.(3)   &  -         \\
0.6 &  0.39(2)    &  4.0   &   7.(1)   &  -         \\
\end{tabular}
\end{ruledtabular}
\end{table}
%%%%%%%%%%%

We define a generalized ``participation ratio'' $\cp_n$
for a single eigenmode via $\cp_n^{-1} = \sum_t (\sum_{\vec{x}}
|\Psi_n(\vec{x},t)|^2)^2$ \cite{DJT}. If $|\Psi_n|^2$ has support mainly on a
four-volume $l_n^4$, then $\cp_n\sim l_n$. A spectral
decomposition of the quantity
$\cp^{-1}(\lambda)= \lim_{m_1\to 0} m_1 \Gamma(t=0;\lambda)$
shows that it is equal to $\rho(\lambda)$ times an average of
$\cp_n^{-1}$ over eigenmodes with eigenvalue $\lambda_n=\lambda$.  Thus, if
we define the ``support length'' $\lsupp = \rho(\lambda)\cp(\lambda)$, we see
that $1/\lsupp$ is an average of $1/l_n$. The fourth column of
Table~\ref{table3} gives this $\lsupp$, a measure of the linear
size of the support of the eigenmodes.

We may now compare $\lsupp$ to the distance between eigenmodes. We
have a measure of the latter from the values of $m_1$ that we used
in measuring $\rho(\lambda)$.  Spectral sums as in
Eq.~(\ref{limm1}) show that $m_1$ is the resolution with which we
detect eigenmodes near $\lambda$.
For  $m_1\simeq 0.01$, the number of modes we detect for a
typical gauge configuration is thus $N \simeq 0.01 V_4 \rho(\lambda)$,
and so $R(\lambda)=(0.01 \,\rho(\lambda))^{-1/4}$ is a
measure of the average distance between modes. If $\lsupp \ll R$
the modes are isolated, and correlation functions reflect
properties of individual localized modes, with no interference.
Table \ref{table3} shows this to be the case well
below the mobility edge.

The density of an exponentially localized mode
has the  asymptotic behavior
\begin{equation}
  |\Psi(x)|^2 \sim c\, \exp(-|x-x_0|/\ltail) ,
\label{ltail}
\end{equation}
which defines $\ltail$, the localization length.
When the modes are isolated, the
decay rate (extrapolated to $m_1=0$) of $\Gamma(t;\lambda)$ reflects the
localization length of the individual modes.
We thus define an average localization
length through $\ltail = 1/\mu$ (Table~\ref{table3}, last
column; compare Table~\ref{table1}, 2nd column). Well below the
mobility edge $\ltail$ turns out to be much smaller than
$\lsupp$. In fact, for $\beta \ge 5.85$, we find that $\ltail <1$ and
$\lsupp>3$. This is good news for domain-wall and overlap
simulations (see below).

Equation~(\ref{ltail}) represents an exponential envelope that we
expect to multiply oscillations in $|\Psi_n|^2$. These
fluctuations, often large, survive the extrapolation of the
correlation function $\Gamma(t;\lambda)$ itself to $m_1=0$. As a result, if
we extract a mass $\tmu$ from $\lim_{m_1\to 0}m_1\Gamma(t;\lambda)$, the result
varies with the details of the fit.  We present
rough values of $\tmu$ in the last column of Table~\ref{table1}.
In the absence of a model for the fluctuations, $\tmu$ is only a
qualitative measure, and hence we do not quote an error.
$\tmu$ follows the trends shown by $\mu$.

Taking all our results together, we have compelling evidence
for exponential localization well below the mobility edge.
Here the modes are isolated in the sense of $\lsupp \ll R$
(for $m_1 \simeq 0.01$), and
$\ltail=1/\mu$ provides an accurate estimate of the average
localization length.
For $\lambda \gtrsim \lambda_c$ interference
effects destroy this connection.
Upon comparing our data for all values of
$(\beta,m_0)$ shown in Table~\ref{table2}, we find that
we can characterize the mobility edge itself as follows:
(1)~$\rho(\lambda)\approx 0.05$--$0.15$ near $\lambda_c$;
(2)~$\ltail\gtrsim1$ signals the proximity of $\lambda_c$; (3)~$\lsupp
\simeq R \simeq 5$ at $\lambda_c$
\cite{future}.

Finally, we revisit the implications of our results for domain-wall and
overlap fermions. These two closely related descendants of Wilson fermions
employ a super-critical Wilson operator as a key element
in their construction. Both are expected to be local, and they both have a
(modified) chiral symmetry \cite{Luscher} at non-zero lattice spacing.
The question is to what extent
these expectations are fulfilled in actual
lattice QCD simulations employing these fermions.  In Ref.~\onlinecite{lcl}
we argued that locality and chirality will coexist in these
formulations if and only if the following holds: On a given
ensemble of configurations, the mobility edge of the underlying
Wilson operator must be well above zero.
It does not matter whether the ensemble is quenched or is generated with
dynamical domain-wall or overlap fermions.
We will not repeat the whole
argument \cite{lcl} here but rather focus on assessing the implications of
our numerical results.

Domain-wall fermions employ an auxiliary, discrete, and (in
practice) finite fifth dimension with spacing $a_5$ and
$N_s$ sites. Finiteness of the fifth dimension ensures locality
but leads to ``residual'' violations of chiral symmetry. A common
measure of these violations, denoted as $\mres$, may be thought of as
an additive correction to the quark mass. It is determined
from a ratio of  pseudoscalar correlation functions at zero spatial
momentum and time separation $t$. In Ref.~\onlinecite{lcl} it was argued
that
\begin{equation}
  \mres \sim c_1 \exp\left(-\tilde\lambda_c N_s\right)
  + c_2 \exp\left[-t\left(\tl_{l}(0)^{-1} - m_\pi\right)\right],
\label{mresx}
\end{equation}
where the two terms arise from extended and localized modes,
respectively.  $m_\pi$ is the pion mass in the simulation.
The mobility edge $\tilde\lambda_c$ and the
localization length $\tl_{l}(\lambda=0)$ refer to a ``hamiltonian'' $\tH$
obtained from the transfer matrix in the fifth dimension, which
depends on $a_5$.
Thus $\mres$ reflects the
spectral properties of $\tH$.
We recover $H_W$ from $\tH$ in the limit $a_5\to 0$.
Moreover, it can be proved that
$\tH\Psi=0$ if and only if $H_W\Psi=0$, i.e. the zero modes of
$\tH$ remain unchanged as $a_5$ is varied.
This implies that $\tl_{l}(0)=\ltail(0)$. For quenched
simulations at $\beta \gtrsim 5.85$ we thus find that
$\tl_{l}(0)=\ltail(0) \simeq 0.6$. The last term in
Eq.~(\ref{mresx}) therefore vanishes rapidly with $t$. This agrees
with previous findings \cite{mres} that $\mres$ is fairly
$t$-independent once $t$ is large enough.
Similarly, $\tilde\lambda_c=0$ if and only if
$\lambda_c=0$. Since good chiral symmetry requires $\mres$ to be
small, it follows that simulations must be performed well
outside the Aoki phase of the underlying Wilson operator.

For overlap fermions, chiral symmetry is guaranteed, but not locality.
Deteriorating locality may distort physical predictions
in an uncontrolled way.
Indeed, for large separations one expects
$D_{\text{ov}}(x,y) \sim c_{\text{ov}} \exp(-|x-y|/l_{\text{ov}})$.
The exponential tail of the overlap may be represented as an unphysical field
of mass $1/l_{\text{ov}}$ that mixes with the physical quarks
with an amplitude controlled by $c_{\text{ov}}$.
The range of the overlap operator, far from being merely a numerical nuisance,
is thus a key indicator of the validity of a simulation.

If an admissibility condition is imposed, it can be proved that
the range $l_{\text{ov}}$ of the overlap operator
is $O(1)$ in lattice units \cite{hjl}. For realistic ensembles,
$l_{\text{ov}}$ depends on the spectral properties of $H_W$,
and good locality again requires keeping away from the Aoki phase.
One anticipates that $l_{\text{ov}}$ is on the order of
either $\lambda_c^{-1}$ or $\ltail(0)$, whichever is larger;
if it is the latter,
$c_{\text{ov}}$ should be related to $\rho(0)$ \cite{lcl}.
The spectral properties studied in this Letter are thus of central
importance for understanding the locality properties of the overlap
operator.

\begin{acknowledgments}
Our computer code is based on the public lattice gauge theory code
of the MILC
collaboration, available from
http://physics.utah.edu/${\scriptstyle\!\sim\,}$detar/milc.html.
We thank the Israel
Inter-University Computation Center for a grant of supercomputer time.
Additional computations were
performed on a Beowulf cluster at SFSU.
This work was supported by the Israel Science Foundation under grant
no.~222/02-1, the Basic Research Fund of Tel Aviv University,
and the US Department of Energy.
\end{acknowledgments}

\end{document}